# Filamentation and self-channeling of spatially modulated high-power femtosecond laser pulses in air

YURY GEINTS,* OLGA MININA, AND ALEXANDER ZEMLYANOV

*V.E. Zuev Institute of Atmospheric Optics, Zuev Square, 1, 634055 Tomsk, Russia*
*Corresponding author: ygeints@iao.ru*

The propagation of high-power femtosecond laser pulses in air under conditions of superposed spatial phase modulation is considered theoretically. The numerical simulations are carried out on the basis of the reduced form of nonlinear Schrödinger equation (NLSE) for time-averaged electric field envelope. Initial spatial modulations are applied to pulse wavefront profiling by a staggered (TEM$_{33}$) phase plate which is simulated numerically. The dynamics of laser pulse self-focusing, filamentation, and post-filamentation self-channeling after the phase plates with variable phase jumps is studied. We show that with specific phase modulations, the pulse filamentation region in air can be markedly shifted further and elongated compared to a non-modulated pulse. Moreover, during the post-filamentation propagation of spatially structured radiation, the highly-localized light channels are formed possessing enhanced intensity and reduced angular divergence which enables post-filamentation pulse self-channeling on the distance multiple exceeding the Rayleigh range.

## 1. INTRODUCTION

High-power ultrashort laser radiation propagates in air, as a rule, in the filamentation mode. It is the result of self-focusing (Kerr) optical nonlinearity of the medium. In the process of laser filamentation, the beam breaks down into tinny optical structures commonly known as the light filaments. They have high intensity and slightly varying transverse width preserving at a considerable long distance [1-3]. Unique nonlinear optical effects accompanying filamentation (generation of higher harmonics, supercontinuum emission, formation of extended plasma channels, generation of terahertz radiation) make it attractive not only for problems of laser technology and materials processing, but also for solving problems of atmospheric optics. As the examples one can consider the creation of extended regions of electrical conductivity in the atmosphere, remote LIDAR diagnostics of the environment, as well as directed transmission of laser energy along extended air paths [4-7].

One of the important research areas on laser radiation self-focusing and filamentation in the atmosphere is the control of the spatial position and structure of the filamentation region. In this regard, it is very promising using specially profiled laser radiation, i.e. laser beams with non-Gaussian transverse intensity distribution. The practical interest in the structured optical radiation is associated with the specific diffraction dynamics of such beams, which in some cases leads to increasing the coordinate of the beam nonlinear focus. This peculiarity also opens up prospects for additional control over the nonlinear propagation area, i.e. the filamentation region. Here, the examples of the advantages of using structured laser radiation are the spatially delayed and extended filamentation of an annular [8] and quasi diffraction-free Bessel-Gaussian beams [9], the multifocal structure of the filamentation area with a superposition of Gaussian and annular profiles (the so-called "dressed" beam) [10, 11], and remote self-focusing of a necklace "corona-beam" [12].

In Ref. [13], the experimental results on air filamentation of femtosecond Ti:Sa laser pulses spatially structured using phase-modulating diffraction elements, a TEM$_{11}$-phase plate (staggered half-wave plate) and a Dammann grating, were recently presented. The authors show that the use of a half-wave plate with checkerboard type spatial phase discontinuities for multifilament regularization may be preferable in two cases. The first one is when the length of the obtained filaments is more important than the stability of the filament configuration as a whole. The second situation is when the very closely spaced multiple filaments are required. In contrast, femtosecond filamentation using a Dammann phase grating does not produce extended filaments, but provides high repeatability and robustness of their spatial configuration (and plasma channels) at a given region on the optical path.

In the present work, we continue the study initiated in [13] and consider a more general case of spatially structured laser pulses filamentation in air. The laser radiation studied has the energy density distribution in the form of separate regularly spaced beamlets of smaller size with the variable but regular wave interphase within a common photonic reservoir. The spatial structuring of the wave phase leading to the following pulse energy partitioning are carried out by phase masks (transparencies) of a checkerboard type with regularly spaced regions with a variable discontinuity of the wave phase at the boundaries of neighboring regions, thus forming a 3×3 phase diffraction grating. The structured radiation formed in this way is then used in our numerical experiments on the nonlinear propagation in air at a relatively long optical distance (several diffraction ranges).

During our simulations we show that the filamentation of the staggered phase-modulated radiation is strongly influenced by the diffraction effects mediated by the Kerr self-focusing which differs from the known filamentation scenario of unimodally profiled laser pulses, e.g., a Gaussian beam. In this case, the diffraction manifests itself already at the spatial scales of individual beamlets. As a result, the filamentation region parameters (the starting coordinate and the length) of the structured pulse can change significantly. Moreover, the use of a staggered phase plate with half-wave phase discontinuities leads to the formation of spatially separated light channels (beams) at the post-filamentation stage with a characteristic intensity of the order of tenths of TW/cm$^2$. An important characteristic feature of these intense channels is their reduced angular divergence in comparison with the divergence of the laser beam as a whole. This opens up the perspectives for using these light channels for various atmospheric applications which demand high-power laser pulse long-range propagation retaining enhanced spatial concentration.

## 2. NUMERICAL MODEL

Consider the nonlinear Schrödinger equation (NLSE) [1] as a basis for numerical simulation of high-power ultrashort laser pulses propagation in a nonlinear medium. Since the solution of this four-dimensional equation requires the use of an extremely huge amount of computational resources, in this work to reduce the computation time, we use its reduced 3D version [14] obtained after integrating the original (3D+1) NLSE over the temporal coordinate. The issues of the admissibility of using such an approach are considered, e.g., in [12]. It is established that such an approximation makes it possible to adequately simulate the main regularities of high-power laser pulses filamentation at moderate propagation distances (several Rayleigh ranges), when the effects associated with the intra-pulse dynamics of the optical field intensity (light shock wave, time front collapse, electron density kinetics) and temporary "memory" of the medium (pulse group-velocity chromatic dispersion, molecular Raman scattering) can be assumed of less importance.

Following [12], we replace the time profile $\tilde{U}$ of a real (Gaussian) laser pulse with a duration $t_p$ by the stepwise Ansatz with a certain duration $t_0$:

$$\tilde{U}(\mathbf{r}_\perp, z, t) = U_0 U(\mathbf{r}_\perp, z), \quad -t_0 \le t \le t_0,$$

$$\tilde{U}(\mathbf{r}_\perp, z, t) = 0, \text{ for all other values of } t. \quad (1)$$

Here, $U(\mathbf{r}_\perp, z)$ is a function of only transverse coordinates $\mathbf{r}_\perp$ and evolutional variable $z$; $U_0 = \sqrt{\pi I_0} t_p / (2 t_0)$ is the amplitude; $I_0$ is the peak intensity of the real pulse, and $\mathbf{r}_\perp$ is the vector of transverse coordinates. The main assumption used in further analysis is that the temporal profile of the laser pulse is considered unchanged during the pulse propagation. This assumption allows to perform the time integration of the full (3D+1) NLSE in advance, neglecting the dynamic effect of the pulse group velocity dispersion, the temporal inertia of the Kerr nonlinearity, and the transient nature of the medium photoionization dynamics. In this case, the model expression for a stepwise pulse provides for significantly simplifying the time integration and allows for obtaining a 3D partial differential equation (PDE) for the average (over time) optical field amplitude: $U(\mathbf{r}_\perp, z) = (1/2T) \int_{2T} \tilde{U}(\mathbf{r}_\perp, z, t) dt$, where $T$ is the boundary value of the temporal numerical grid. The parameter of probe pulse duration $t_0$ is free. Its value is usually chosen in the range $0 < t_0 \le t_p$, provided that the best agreement with the results of calculations of the full (3D + 1) model of NLSE is obtained. In the case considered here, the value $t_0 = 0.15 t_p$ is adopted, which qualitatively takes into account the characteristic temporal compression of the pulse due to Kerr self-focusing.

First of all, we consider the rate equation for the instantaneous plasma density of free electrons $\rho_e$:

$$\frac{\partial \rho_e}{\partial t} = W_I(I)(\rho_{nt} - \rho_e) + \nu_i \rho_e I - \nu_r \rho_e^2, \quad (2)$$

where $W_I$ is the photoionization rate, $I = |U|^2$ is pulse intensity, $\rho_{nt}$ is the density of neutral atoms (molecules), $\nu_i$ is the rate of impact ionization. All losses of $\rho_e$ (recombination, attachment, and free diffusion) are absorbed by the term $(\nu_r \rho_e^2)$ on the right-hand side of Eq. (2). Using the Eq. (1) and integrating the rate equation with respect to time, we obtain the solution to Eq. (2) for normalized plasma electron density $\bar{\rho}_e = \rho_e / \rho_{nt}$ as follows:

$$\bar{\rho}_e(\bar{t}) = \frac{R \tanh(0.5 R t_0 [\bar{t} + C]) - [W_I - \nu_i |U|^2]}{2 \nu_r \rho_{nt}}, \quad (3)$$

where $\bar{\rho}_e = \rho_e / \rho_{nt}$, $R = \sqrt{[W_I - \nu_i |U|^2]^2 + 4 W_I \nu_r \rho_{nt}}$, $C = (2/R t_0) \tanh^{-1}([W_I - \nu_i |U|^2]/R)$ is the constant of integration, determined from the condition $\bar{\rho}_e(\bar{t} = 0) = 0$.

Now, within the approximations made we obtain the following three-dimensional NLSE for the time-averaged electric field amplitude $U(\mathbf{r}_\perp, z)$:

$$\left\{ \frac{\partial}{\partial z} - \frac{i}{4} \nabla_\perp^2 - i \frac{L_R}{L_K} |U|^2 + \frac{L_R}{2 L_W} W_I \frac{(1-B)}{|U|^2} + \frac{i L_R}{2 L_{pl}} \left(1 - \frac{i}{\omega_0 \tau_c}\right) B(U) \right\} U = 0, \quad (4)$$

where $\nabla_\perp^2 = \frac{\partial^2}{\partial x^2} + \frac{\partial^2}{\partial y^2}$ is the transverse Laplacian, $\omega_0 = 2\pi c/\lambda_0$ is the central pulse angular frequency, $\tau_c$ is the characteristic electron collision time. This equation includes the following characteristic parameters of the problem solved: (a) beam Rayleigh range, $L_R = k_0 R_0^2 / 2$ ($L_R = 4$ m for a Gaussian beam with $R_0 = 1$ mm and the wavelength 800 nm), (b) Kerr self-focusing length, $L_K = n_0/(k_0 n_2 I)$, (c) plasma refraction length, $L_{pl} = n_0 \Delta E_i / (2 \sigma_c \nu^{(K)} I^K t_p \omega_0 \tau_c)$, as well as (d) the length for the multiphoton ionization, $L_W = (\nu^{(K)} I^{K-1})^{-1}$, where $\nu^{(K)}$ is the multiphoton ionization rate of a molecule provided by simultaneous absorption of $K$ photons of laser radiation. Here, $k_0$ is the wavenumber, $n_0$ is the refractive index of the unperturbed medium, $n_2$ is the refractive index cubic nonlinearity $n = n_0 + n_2 I$, $\sigma_c$ and $\Delta E_i$ are cascade ionization cross-section and the ionization potential of the molecule, respectively. The functional $B = B(W_I, \nu_i, |U|^2)$ takes into account the total (over the pulse) gain of free electrons in the beam wake upon multiphoton ($W_I$) and impact ($\nu_i$) ionization of gas molecules thus providing the saturation of the self-focusing (Kerr) nonlinearity and plasma absorption:

$$B(U) = \frac{\ln\left\{\frac{e^{t_0(RC+2R)}+1}{e^{t_0 RC}+1}\right\} - t_0 [R + (W_I - \nu_r |U|^2)]}{2 t_0 \nu_r \rho_{nt}}. \quad (5)$$

The multiphoton photoionization rate $W_I$ is calculated using the Perelomov-Popov-Terent'ev model [15], which is applied to a gas mixture simulating atmospheric air (20%$O_2$+80%$N_2$). The total neutral molecule concentration is 2.5·$10^{19}$ $cm^{-3}$. The critical self-focusing power $P_{cr}$ for a femtosecond laser pulse at the wavelength of $\lambda$ = 800 nm in air is set as 3.2 GW [16]. Additionally, in Eq. (4) the linear air absorption of laser radiation is neglected because of the smallness of corresponding absorption coefficient in the near-infrared spectral band (< $10^{-6}$ $m^{-1}$).

In the numerical calculations, the initial transverse distribution of the laser radiation $U(\mathbf{r}_\perp, z = 0)$ possesses a plane wavefront and is set in Descartes coordinates $\mathbf{r}_\perp \equiv (x, y)$. Additionally, the realistic beam quality degradation inherent to powerful laser sources is simulated by superimposing a random amplitude noise on the generated smooth intensity profile as follows: $U_\perp(x,y) \rightarrow U_\perp(x,y)\left[1 + A_m \cdot \tilde{f}(x,y)\right]$, where $\tilde{f}$ is a random variable normally distributed in the range [-1…1], and $A_m = U_0/20$ is the noise amplitude. To be specific, all calculations are performed for laser beams with the initial radius (at the 1/e level from the maximum intensity) $R_0$ = 1 mm and pulse duration $t_p$ = 50 fs. The peak pulse power $P_0$ is varied from 3 to 25 critical powers of self-focusing $P_{cr}$.

The propagation of high-power femtosecond laser pulses in air is simulated under conditions of additional modulation of the wave phase, which is implemented by phase transparencies consisting of 9 quadratic segments arranged in a staggered form. These segments provided a set of the wave phase discontinuity $\Delta\varphi$ at the boundaries of adjacent segments in the range from -$2\pi$ to $2\pi$. Examples of the considered phase-modulating transparencies for the phase jump $\Delta\varphi$ = -$\pi$ and $\pi$ are shown in Figs. 1(a-c). As will be seen below, the use of such transparencies ensures the partitioning the initial unimodal laser beam into the separate parts, which allows for the minimization the effect of small-scale self-focusing and controlling the characteristics of the multiple filamentation area and post-filamentation light channels (PFC). Worth noting, the features of post-filamentation propagation of femtosecond laser pulses are studied in detail in [17-19].

One of the additional parameters that can be varied during the simulation is the presence (Figs. 1(a),(c)) or absence (Figs. 1(b),(d)) of the central element modulation in the phase matrix. The last case produces a beam with the unperturbed most intense central part. As will be seen below, such phase masks provide an additional displacement of the filamentation beginning for practically all considered configurations of phase-modulating transparencies. For the convenience of further analysis, we enumerate the phase transparencies with Latin numerals from I to IV in accordance with the order shown in Fig. 1.

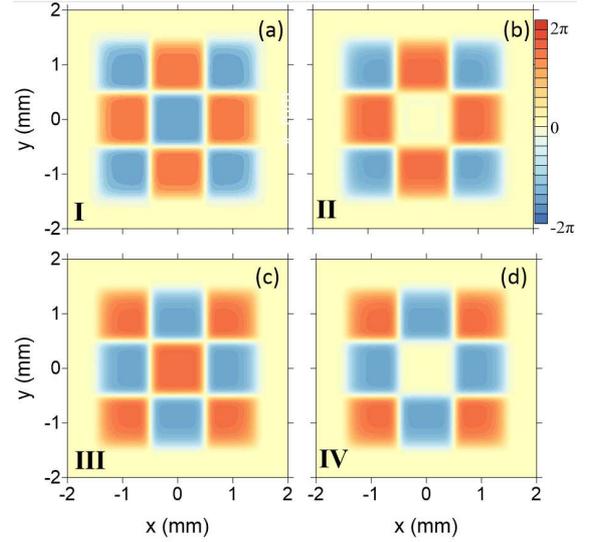

Fig. 1. Phase modulating transparencies used in simulations with the phase jump between adjacent segments (a, b) $\Delta\varphi$ = -$\pi$ and (c, d) $\Delta\varphi$ = $\pi$ in cases with (a, c) modulated and (b, d) unmodulated central element.

Before proceeding to the results of our simulations, we present the benchmarking of full and time-integrated optical models for one of the phase masks shown in Fig. 1 in order to confirm the validity of the reduced 3D NLSE used. Test calculations of the nonlinear propagation of a laser pulse in air are performed within the full model described by the spectral version of the vectorial wave equation for the complex electric field envelope, $U_{k\omega} \equiv U_\perp(k_\perp, z; \omega)$, in the spatiotemporal frequency range which is known in the literature as the unidirectional pulse propagation equation (UPPE) [20, 21]. To this end, the following equation is solved numerically:

$$\frac{\partial U_{k\omega}}{\partial z} = i(k_z - \omega_0/v_g)U_{k\omega} + i\frac{\omega^2}{2c^2 k_z}\frac{P_{k\omega}}{\varepsilon_0}. \quad (6)$$

Here, $k_z = \sqrt{k(\omega)^2 - k_\perp^2}$ is the propagation constant along $z$, $k_\perp^2$ is the squared transverse wavevector component, $k(\omega) = \omega n(\omega)/c$ is the wavenumber depending on the pulse angular frequency $\omega$ and the chromatic dispersion of air $n(\omega)$ that is modelled by the Sellmeier equation, $\varepsilon_0$, $c$ are dielectric permittivity and light speed in vacuum, respectively, $P_{k\omega}$ is the nonlinear medium polarization which accounts for all significant physical effects causing pulse self-phase modulation and nonlinear energy dissipation during the propagation [20]. Eq. (6) is formulated in the pulse coordinate frame with the origin moving with pulse group velocity $v_g = (\partial k/\partial\omega)^{-1}$.

As an example, Figs. 2(a, b) show the dynamics of peak pulse intensity Imax and the density of free electrons $\rho_e$ of laser plasma realized in air when using the phase mask IV. As seen, on the whole the reduced NLSE satisfactorily describes the nonlinear propagation of a femtosecond laser pulse under conditions of beam phase modulation. This concerns both the peak radiation parameters and the spatial scales of the filamentation area. In the case shown, the mean computation time of the model for the same initial parameters of laser pulse is reduced by almost an order of magnitude and not exceeds a couple of hours when using the reduced NLSE. It should be noted that a fairly good agreement between both models preserves even for larger pulse peak power ($\eta = P_0 / P_{cr} = 20$). From the analysis of Fig. 2 it follows that the main differences in the results obtained within the two models lay in the spatial characteristics of the filamentation area, in particular, in the beginning and ending coordinates, as well as in the filamentation longitudinal continuity. UPPE calculations exhibit an earlier start of a phase-modulated pulse filamentation and earlier filaments decaying as compared with the reduced NLSE model. However, the total length of the filamentation region is approximately the same in both propagation models. Similar simulations performed for laser pulses with a lower peak power (not shown here) demonstrate an even better agreement between the two propagation models.

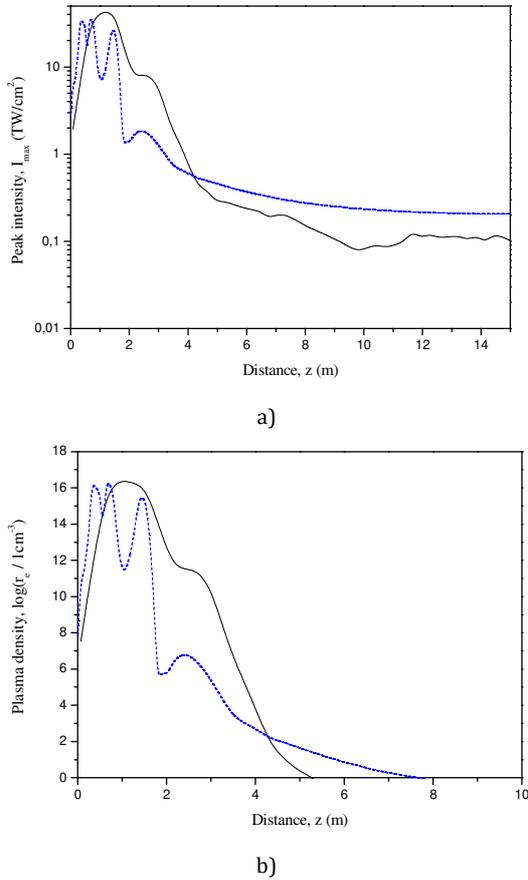

Fig. 2. Dependence of (a) peak intensity $I_{max}$ and (b) maximum plasma density $\log(\rho_e/1[cm^{-3}])$ on the propagation distance for laser pulse with relative peak power $\eta = 20$ modulated by the phase mask IV ($\Delta\varphi = \pi$, central element is off). Solid curves show the full (3D+1) UPPE model, while dashed curves represent the reduced 3D-NLSE.

## 3. RESULTS AND DISCUSSION

The results of our numerical simulations of laser pulse filamentation with different phase-modulation transparencies are shown in Figs. 3(a, b) for two relative pulse power, $\eta = 5$ and 15. Hereafter, it is instructive to express the phase jump of the phase plate $\Delta\varphi$ in units of $\pi$ introducing the dimensionless value $\varphi_0 = \Delta\varphi/\pi$. Recall that $\Delta\varphi$ is the phase jump between the neighboring transparency segments starting from the upper left element (see, Fig. 1). From the analysis of Fig. 3 it follows that generally the use of phase mask leads to a reduction in the self-focusing distance of the femtosecond laser pulse in comparison with an unmodulated Gaussian beam. This reduction increases when the phase shift $\varphi_0$ of the transparency segments becomes higher. Note, the boundaries of the pulse filamentation region are determined from the outermost spatial coordinates where the peak plasma density $\rho_e$ equals to the specific value of $10^{14}$ cm$^{-3}$.

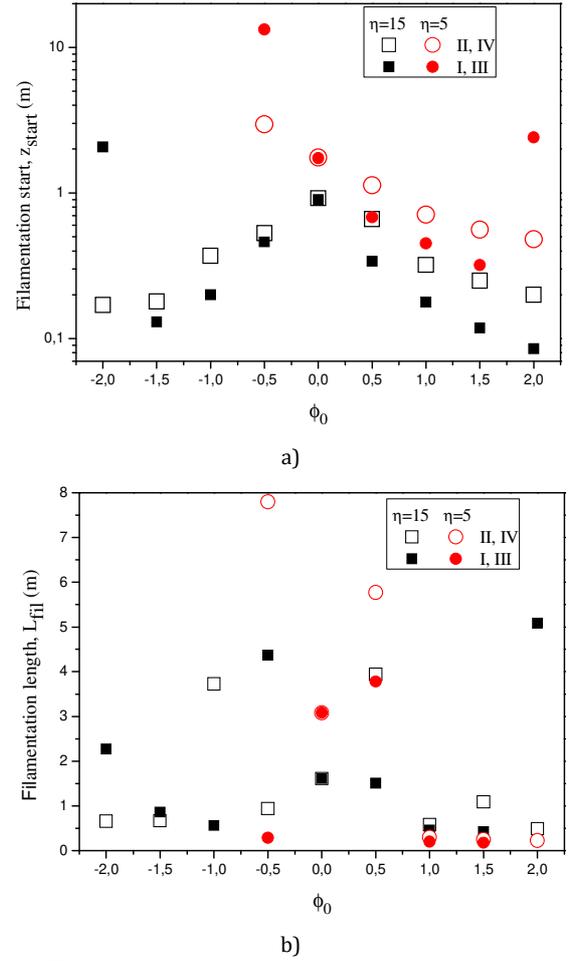

Fig. 3. The filamentation (a) starting distance $z_{start}$ and (b) length $L_{fil}$ for different phase-modulating masks and pulse relative power η.

However, for particular phase plates considered here the start of pulse filamentation occurs at larger distances $z_{start}$ from the optical path beginning as compared to the case when the phase modulation transparency is not used ($\varphi_0 = 0$). In Fig. 3(a), these situations correspond to the values $\varphi_0 = $ -2, -0.5, and 2. Importantly, at $\varphi_0 = $ -0.5 a significant (several times) increase in the length $L_{fil}$ of the filamentation area is also observed in Fig. 3(b). In this case, the use of a phase-modulating transparency may not only shift the coordinate of the filamentation beginning (its decrease or increase, as, e.g., at $\eta = 5$ and $\varphi_0 = $ -0.5 in Fig. 3(a)), but also may completely prevent the filamentation even for supercritical pulse power as, e.g., at $\eta = 5$ and $\varphi_0 < -1$. In the last case, strong divergence of the entire beam, as well as the individual spatial pulse parts (beamlets) is observed already at the initial stage of pulse propagation. The result of this aberrational defocusing is that the maximum pulse intensity does not reach the values characteristic for the femtosecond filamentation (> 30 TW/cm$^2$), and turns out to be an order of magnitude lower. The concentration of free electrons produced by the pulse in this case does not exceed $10^{10}$ cm$^{-3}$. Nevertheless, even for such situations the laser beam partitioning by a staggered phase modulating plate makes it possible obtaining several extended light channels with a moderately high intensity and low angular divergence (see below).

As an example, let us consider in more detail the propagation of a femtosecond laser pulse with a relative peak power $\eta = 15$ passed through the transparency with $\varphi_0 = -1$ (wave phase jump $\Delta\varphi = \pi$, transparency I in Fig. 1(a)). Peak pulse intensity and the laser plasma density in the beam wake for this case are shown in Fig. 4(a). 2D surface maps in Figs. 4(b-g) show the transverse fluence profiles of laser pulse at different propagation distances in air. Note, in each figure the fluence distribution is normalized to its maximum.

Fig. 4(b) shows that the pulse filamentation starts at $z = 0.5$ m in the beam center at the very initial stage of the propagation. A clear manifestation of optical diffraction at sharp phase gradients is also evidenced by the appearing of the "craters" located in the centers of the side lobes in the fluence distribution. During the propagation, the optical energy density in the beam center decreases and the pulse filamentation stops, but at some distance ($z = 1.5$ m in Fig. 4(a)) the side lobes begin to increase again due to focusing of the beam periphery (Fig. 4(c)). In this case, the free electron density in the beam wake also increases, but this rise turns out to be insufficient for pulse refilamentation ($\rho_e < 10^{10}$ cm$^{-3}$). As a result, from this distance on the laser pulse enters the stage of post-filamentation propagation in the form of five isolated high-intensity light channels (Fig. 4(d)). The formation of these post-filamentation channels (PFCs) occurs independently in the opposite beam sections encircling the center and is clearly observed in Fig. 4(d). PFCs continue to exist along the remaining optical path but their number first increases up to nine (Fig. 4(f)) and then drops again to five at the end of the propagation in Fig. 4(g). Note, in our simulations a PFC is determined from the condition that a corresponding lobe in 2D fluence distribution should possess a maximal value greater or equal to half of the central maximum.

A characteristic feature of the propagation of such spatially structured pulse is also the presence of a circular ring in the fluence distribution (depicted by a bold circle in Figs. 4(f) and (g)) encompassing the central maximum. This ring structure occurs due to the constructive interference of optical radiation coming from the peripheral and central segments of the phase modulating transparency. Worthwhile, the appearance of such a ring structure contributes to the maintenance of low angular divergence of the most intense central PFC. Importantly, the slowly decreasing pulse intensity in PFC remains at the subterawatt value ($I_{max} > 0.1$ TW/cm$^2$) even at distances greater than multiple Rayleigh range ($z \approx 2L_R$).

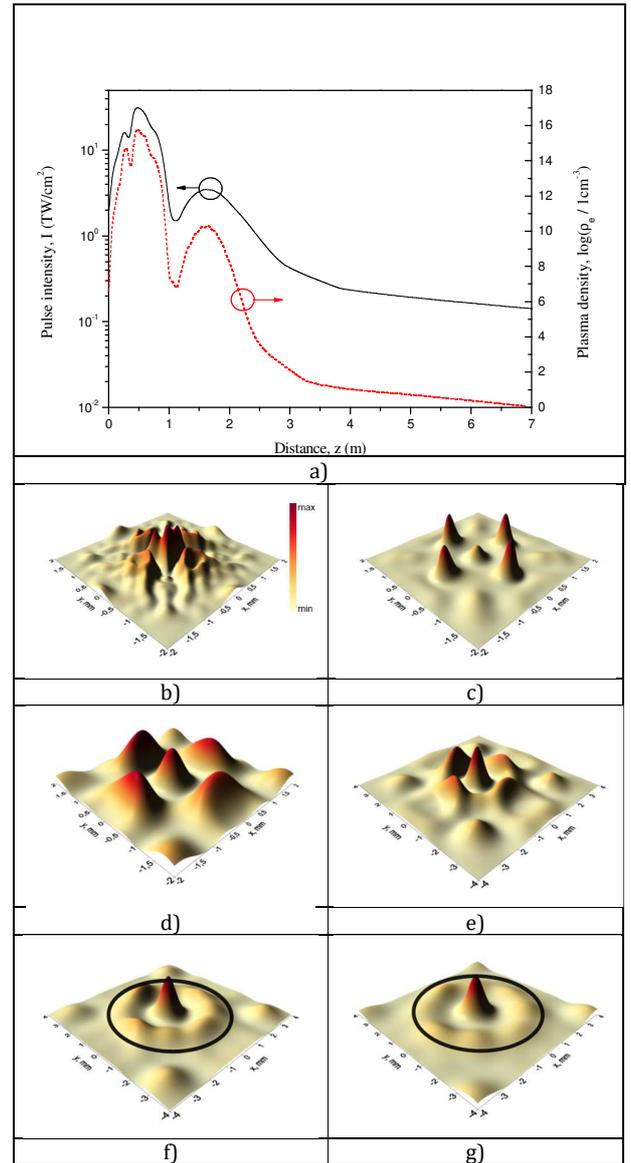

Fig. 4. (a) Maximum intensity and free electrons concentration along the air path during the propagation of a laser pulse with $\eta = 15$ modulated by phase plate I with $\varphi_0 = -1$. (b-g) Pulse cross-sections taken sequentially at the distances from 0.5 to 5.5 m with the step of 1 m.

Similar dependences in the formation and propagation of the PFCs are observed for other values of pulse power when using phase masks with $\varphi_0 = -1$ and $+1$, as presented in Figs. 5(a, b) for the longitudinal profile of pulse peak intensity. In these cases, the peculiarity of laser pulse structuring ensures the quasi-independent dynamics of the adjacent beamlets propagation separated by neighboring phase plate segments. Here, a specific pulse propagation regime can be realized featuring a weakly varying intensity at the long-range post-filamentation stage where the self-channeling of the PFCs takes place. This propagation regime is very promising for the problems of long-range propagation of high-power femtosecond laser pulses. A striking example of long-range PFC self-channeling is shown by green dashed line in Fig. 5(b) for a laser pulse with the relative power $\eta = 20$ modulated by the phase transparency IV (see,

Fig. 1(d)). In this case, the peak intensity values of approximately 0.1 TW/cm² are maintained at the distance of about $3L_R$.

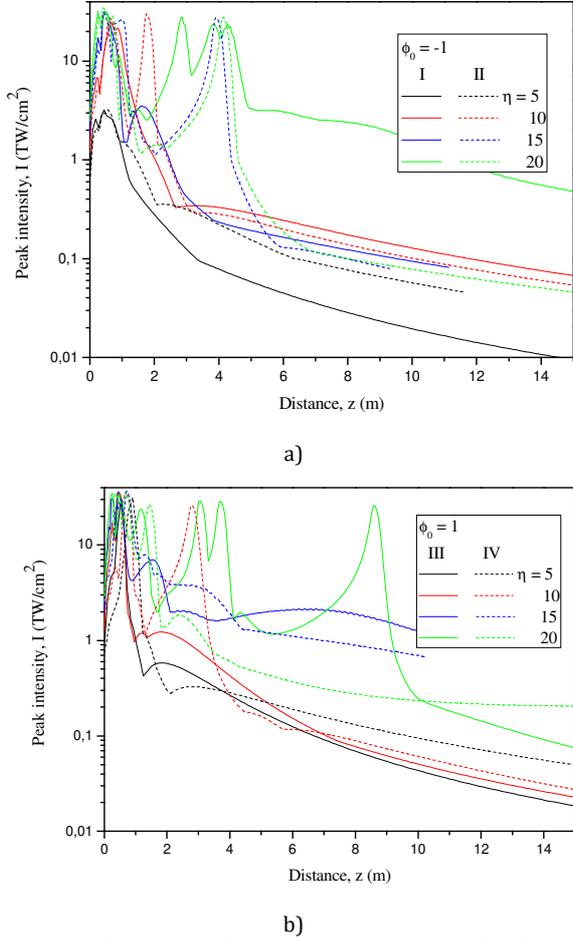

a)

b)

Fig. 5. Peak intensity along pulse propagation with different relative power η for phase transparencies with $\varphi_0$ = -1 (a) and 1(b).

Now consider another important parameter characterizing the propagation dynamics of the structured optical radiation using phase-modulating transparencies. This parameter is the angular divergence of optical beam at the stage of post-filamentation propagation when the beam takes the form of one or several light channels located at the nodes of a symmetric spatial grating as in Figs. 4(c)-(f). We calculate the angular divergence $\theta$ of every PFC from the dependence of its radius $R$ with increasing propagation distance $z$ according to the standard expression: $\theta = 1/2\sqrt{\partial^2 R/\partial z^2}$. This is shown in Fig. 6(a) for the value of relative divergence $\theta/\theta_0$ while Fig. 6(b) presents the corresponding $R(z)$ dependence. Here, $\theta_0$ determines the PFC angular divergence when no phase modulating transparency is employed ($\varphi_0$ = 0).

Generally, the angular PFC divergence decreases with pulse power increasing that is consistent with previously published results [22]. This tendency is observed regardless of the phase jump value $\varphi_0$ and the type of the phase plate used (presence or absence of the central element modulation). Worth noting, in some situations the use of a phase-modulating transparency allows for the obtaining high-intensity light channels with lower divergence than that obtained without pulse spatial modulation, thus $\theta < \theta_0$. For example, one can consider the case obtained using the phase plate with $\varphi_0$ = -0.5 and pulse power $\eta$ = 9, as well as with $\varphi_0$ = -1 and $\eta$ = 20 (transparency I) where the ratio $\theta/\theta_0$ does not exceed 0.4. In these cases, the formation of a moderate number of well spatially separated PFCs (no more than 6 channels) is observed, which surround the central high-intensity part of the beam. By analogy with the filamentation dynamics of a "dressed" beam [11, 23], one can claim that a part of the pulse energy contained in the annular area of PFCs serves as a source of additional energy replenishment of the central light channel and provides for this PFC elongation. Moreover, the circularly-distributed PFCs in addition to performing a purely energy function of feeding the center forms a specific "diffractive waveguide" in the propagation medium [8], which also assists to angular divergence reducing of the central PFC.

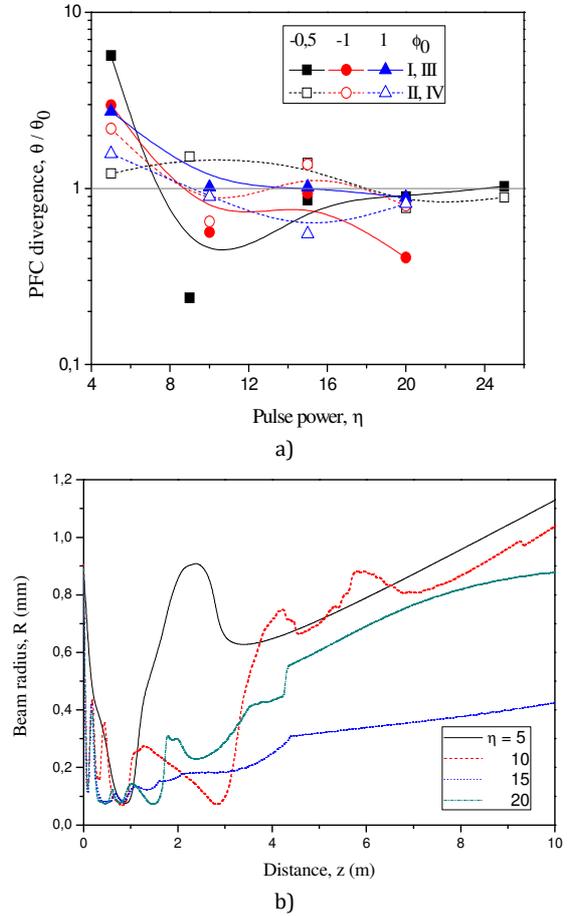

a)

b)

Fig. 6. (a) Dependence of the relative PFC divergence $\theta/\theta_0$ on relative pulse power $\eta$. (b) PFC radius along the propagation path for pulses with different relative peak power $\eta$ and a phase-modulating transparency IV with $\varphi_0$ = 1.

The number $N_c$ of separate PFCs which are clearly distinguishable in the fluence profiles to the end of pulse propagation is presented in Fig. 7 for different pulse power and phase transparencies used. For clearness, here we consider only three values of the phase jump $\varphi_0$. Other $\varphi_0$ values exhibit similar dependencies. As seen, the channel number $N_c$ for different $\eta$- values and the type of phase-modulating transparency can vary from 1 to 9. We can note three most typical cases of PFC arrangement: (a) the presence of a single central PFC, (b) the formation from 8 to 9 PFCs by every "chessboard" element of the phase transparency, and (c) an intermediate case when there is a moderate, from 5 to 6 light channels. In general, the phase plates of II and IV types (without central element modulation) give larger number of the post-filamentation channels regardless the pulse power.

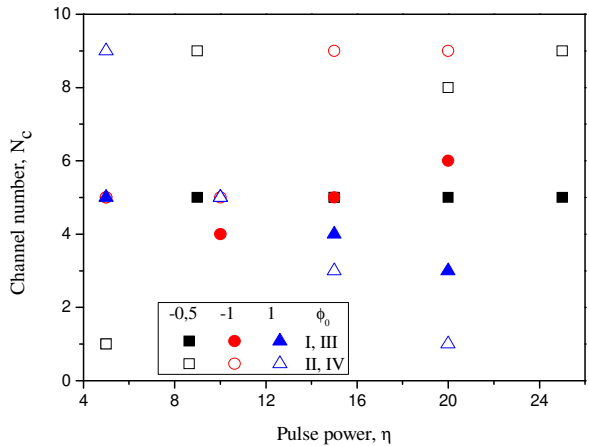

Fig. 7. The final PFC number *versus* relative pulse power for phase transparencies with $\varphi_0$ = -0.5, -1 and 1.

## 4. CONCLUSIONS

In conclusion, the filamentation dynamics of high-power femtosecond laser pulse propagation in air is considered under the conditions of applied spatial modulation by staggered $TEM_{33}$-type phase plates with variable phase jumps between adjacent elements. On the basis of numerical solution to the time-averaged nonlinear Schrödinger equation, the computer simulations are carried out for various initial configurations of phase-modulating transparencies and initial pulse power. We found that two phase plates with a $\pi/2$ phase jump differing by the presence or absence of the phase modulation in the central element provide for the filamentation region shift far away from the optical path beginning and for increasing the filamentation length. Meanwhile, such spatial structuring of an optical pulse results in its breaking into several well-separated high-intensity light channels at the stage of post-filamentation propagation. Generally, this propagation stage conditionally begins at the distances of the Rayleigh range in the cases under consideration. Normally, the angular divergence of these high-intensity PFCs does not exceed the angular divergence of the channel formed when no phase-modulating transparency is used except the case of low pulse power. However, in the situations with $-\pi/2$ and $-\pi$ phase jumps and moderate pulse power, the use of a phase-modulating transparency allows for the obtaining reduced PFC angular divergence.

**Funding.** This work was supported by the Russian Science Foundation (Agreement № 21-12-00109).